\documentclass[%
reprint,
superscriptaddress,
nofootinbib,
 amsmath,
 amssymb,
 aps,
 prc,
]{revtex4-2}

\usepackage{hyperref}
\usepackage{amsmath}
\usepackage{amssymb}
\usepackage[dvipdfmx]{graphicx}
\usepackage{mathrsfs}
\usepackage{color}
\usepackage{bm}
\usepackage{bbm}

\begin{document}

\title{Visualization of nuclear many-body correlations with the most probable configuration of nucleons}

\author{Moemi Matsumoto}
\affiliation{Department of Physics, Tohoku University, Sendai, 980-8578, Japan}

\author{Yusuke Tanimura}
\affiliation{Department of Physics, Tohoku University, Sendai, 980-8578, Japan}
\affiliation{Graduate Program on Physics for the Universe, Tohoku University, Sendai, 980-8578, Japan}

\date{\today}

\begin{abstract}

A method to visualize many-body correlations using the information of 
the full wave function is presented. 
The set of nucleon coordinates which maximizes the square of the wave function, 
that is, the most probable spatial configuration of nucleons, is visualized. 
The method is applied to Hartree-Fock (HF) and HF+BCS wave functions of $p$- and $sd$-shell $N=Z$ 
even-even nuclei to analyze the many-body correlations in those systems. 
It is found that there are $\alpha$-cluster-like four-body correlations already at the HF level in some of the nuclei. The effects of pairing on the most probable configuration are also investigated. 
The method is useful to analyze the nuclear many-body correlations, and 
it suggests a new viewpoint to microscopic nuclear wave functions. 
\end{abstract}


\maketitle

\section{Introduction}
The nuclear many-body wave function contains an enormous amount of information 
since it is, for an $A$-body system, a function of $3A$ continuous 
variables for positions and $2A$ discrete variables for spins and isospins.  
Therefore, in general it is difficult to analyze the full information of the wave function. 
Moreover, many of the experimental observables are expectation values or transition 
probabilities of one- or two-body operators. For these reasons, theoretical analyses in 
nuclear physics are made mainly for quantities obtained after integrating out most of 
the information of correlations embedded in a many-body wave function, such as one- and two-body densities. 

A typical example of spatial nucleon correlations is clustering. 
Cluster model calculations have been successfully applied mainly to light systems where some specific cluster structure is explicitly assumed \cite{OFK06,B1,B2,ZRM94}. 
Studies of the cluster structure based on microscopic theories without assuming 
any clusters have been made with the one-body density \cite{GS13, Eb14,Fr18,KKO12} and 
two-body density (the localization function) \cite{RMUO11,ZSN16,SN17,EKN17,Ta19,Ren22} 
in the coordinate space. 
There was also an analysis of the overlaps between $\alpha$-cluster-model and mean-field wave functions \cite{Ma06}.
The reduced width amplitude and spectroscopic factor can also be used to 
quantify the existence probability of clusters \cite{Ho77,KH06,KEST14,Ta21}. 

In the field of quantum chemistry, on the other hand, methods to visualize the 
spatial correlation among all electrons in a system have been developed and applied to 
studies of molecular structures \cite{Liu16,Liu20_NC,Liu20_JPCL,Liu19}. 
The authors of Ref. \cite{Liu16} have developed a method, which they call 
``dynamic Voronoi metropolis sampling (DVMS)'', to efficiently compute the average 
electron configuration of the system by partitioning the $3N$-dimensional space 
of a many-electron wave function into regions related by the permutation symmetry. 
It provides a way to find out a ``representative snapshot'' of all the electrons 
in a molecule. 
The method does not rely on the molecular orbitals, which cannot be defined uniquely 
for a Slater determinant, and are even less well defined when the wave function is given by a superposition 
of many Slater determinants. Thus it gives a robust way to analyze the wave functions 
obtained by any theoretical framework, and it would be interesting to apply such a 
method to the nuclear structure as well.

In this paper, we employ a simpler method: the most probable spatial arrangement of nucleons, {\it i.e.}, the set of coordinates 
$(\bm r_1\sigma_1\tau_1,\bm r_2\sigma_2\tau_2,\dots,\bm r_A\sigma_A\tau_A)$ that maximizes 
$|\Psi|^2$, is searched for and visualized as the likely snapshot of a nucleus. 
As the first application of the method, 
we test it with nuclear many-body wave functions obtained by Hartree-Fock (HF) and HF+BCS theories. 

The paper is organized as follows. 
In Sec. \ref{sec:method}, the method to find the maximum of $|\Psi|^2$ is introduced. 
In Sec. \ref{sec:results} we describe the results for the $p$- and $sd$-shell 
$N=Z$ nuclei. A summary and future outlook are given in Sec. \ref{sec:summary}. 
In addition, we give in Appendix \ref{app:bcs} the coordinate-space representation of the BCS-type wave function.

\section{Method: $|\Psi|^2$ maximization}\label{sec:method}

In this section, we present the method that we call the ``$|\Psi|^2$-maximization method'' for visualization of the many-body correlations. 
The idea is to plot the most probable configuration of all the nucleons in a nucleus. 

We consider a time-even many-body state consisting of $N (={\rm even})$ identical fermions. For simplicity we ignore the isospin degrees of freedom. 
The function to be maximized is the square of the wave function, 
\begin{eqnarray}
\rho^{(N)}(x_1,x_2,\dots,x_N) &\equiv& |\Psi(x_1,x_2,\dots,x_N)|^2, 
\label{eq:rhoN}
\end{eqnarray}
where $x_i \equiv (\bm r_i\sigma_i)$ denotes the position and spin variables of the $i$-th particle. 
The superscript $(N)$ emphasizes that this is the ``$N$-body density'', 
which is the probability density of finding the $N$ particles simultaneously at the given set of coordinates. 
Notice that $\rho^{(N)}$ is invariant under any permutation of the coordinates. 
To simplify the problem, we define the density within a limited domain of the parameter space, 
\begin{eqnarray}
&&
\rho_{d_s}^{(N)}(\bm r_1,\bm r_2,\dots,\bm r_N)
\nonumber\\
&\equiv&
\rho(\underbrace{\bm r_1\uparrow,\ldots,\bm r_{N/2+d_s}\uparrow}_{N/2+d_s\ {\rm up}},
\underbrace{\bm r_{N/2+d_s+1}\downarrow,\ldots,\bm r_{N}\downarrow}_{N/2-d_s\ {\rm down}}), 
\label{eq:rhoN_ds}
\end{eqnarray}
with the spin variables fixed. 
The integer parameter $d_s$ denotes half the difference between the numbers of spin ups 
and downs. 
For example, $d_s=0$ gives
\begin{eqnarray}
&&
\rho_{0}^{(N)}(\bm r_1,\bm r_2,\dots,\bm r_N)
\nonumber\\
&\equiv&
\rho(\underbrace{\bm r_1\uparrow,\ldots,\bm r_{N/2}\uparrow}_{N/2\ {\rm up}},
\underbrace{\bm r_{N/2+1}\downarrow,\ldots,\bm r_{N}\downarrow}_{N/2\ {\rm down}}), 
\end{eqnarray}
for which the numbers of spin ups and downs are equal. 
The function $\rho_{d_s}^{(N)}$ depends only on continuous variables $\bm r_1,\bm r_2,\dots,\bm r_N$. Note again that the order of the spatial coordinates within those with the same 
spin orientation is irrelevant. 
Due to the time-reversal invariance of the many-body state, it suffices to consider 
only $\rho_{d_s\geq 0}^{(N)}$. Thus the task is to maximize $\rho_{d_s}^{(N)}$ for 
$0\leq d_s\leq N/2$ and locate the global maximum. 
Then we plot the set of coordinates, which maximizes $\rho^{(N)}$, in the three-dimensional 
space, specifying also the spin orientations. 

The $|\Psi|^2$-maximization method gives a qualitative and intuitive 
picture for the many-body correlations but does not give a quantitative 
measure for degree of clustering or other types of correlation. 
The method is still under development towards more extensions and applications as we will mention in Sec. \ref{sec:summary}.

Note that, by maximization, one only finds the maximum of the probability distribution 
and does not pay attention to its global behavior, such as the fluctuation around the 
maximum or the existence of local maxima. It is also pointed out in Ref. \cite{Liu16} that 
the maximum of an electronic wave function in a molecule is not always representative, 
and it may lead to a misleading picture for molecular-bond structure. 
Therefore, to study the many-body correlations in more detail, 
it is important to investigate the global behaviors as well as the maximum. 
Nevertheless, in this paper, we try this simple $|\Psi|^2$ maximization to investigate 
what we can see from the nuclear many-body wave function. 

We should also point out that analyses of the full wave function $|\Psi(x_1,\dots,x_N)|^2$ 
or its maximum have already been made in few-body calculations \cite{Zh94,HS05,HSCS07,Hi19}. 
There is also a method to determine the ``physical coordinates'' of nucleons in the
antisymmetrized molecular dynamics wave function \cite{Ono92}. 
The $|\Psi|^2$-maximization presented in this work gives a more general way to 
perform similar analyses for systems of many particles and for general types of wave functions. 

\section{Results and discussion}\label{sec:results}

We test the $|\Psi|^2$-maximization method with 
nuclear wave functions of light $N=Z$ nuclei obtained by microscopic theories. 

Due to the short-range and attractive natures of the nucleon-nucleon force, 
it is reasonable to assume for the nuclear ground states that the maximum of $\rho_{d_s=0}^{(N)}$ 
is the global maximum of $\rho^{(N)}$. 
Thus we will show in this section only the maximum of $\rho^{(N)}_0$ for all the systems that we examine 
(see Appendix \ref{app:ds} for $d_s$ dependence of the maximum value of $\rho_{d_s}^{(N)}$ 
for $^{20}$Ne nucleus as an example). 
The neutron-proton formalism is employed in the present framework, and 
the half of the spins are fixed to be up and the other half down each 
for neutron and proton sector. 

\subsection{Set up}

We take Hartree-Fock (HF) and HF+BCS wave functions 
with the SLy4 Skyrme effective interaction \cite{CBH98} for the HF part
and the constant-gap approximation for the BCS part. 
The HF+BCS state is obtained as usual by adjusting the Fermi energies to get the 
correct average particle numbers. 
We impose the axial and reflection symmetries, and the time-reversal symmetry \cite{Va73}. 
Note that, in the present calculation, the wave function is given as a product of 
the neutron and proton parts, each of which depends only on the neutron and proton 
coordinates, respectively. 
Therefore, there is no explicit correlation between neutrons and protons, 
and the maximum search can be carried out separately. 
The maximum search is performed with the conjugate gradient (CG) method \cite{NR}
starting from a random initial configuration for the set of spatial coordinates. 

One needs values of the wave function $\Psi(x_1,x_2,\dots,x_N)$ for any given set of coordinates. 
The wave function of a HF state is given by a Slater determinant.
The wave function of the $N$-particle component of a HF+BCS state is 
given by a Pfaffian. See Appendix \ref{app:bcs} for the explicit expression of the 
BCS wave function in the coordinate-space representation.  

Some remarks on the correlations present in the mean-field wave function are in order. 
The Pauli principle, of course, is explicitly taken into account by antisymmetrization. 
There are also the long-range correlations through the mean fields due to the interaction, 
which roughly determine the nuclear (intrinsic) density distribution \cite{RS}. 
In particular, the deformation induces collective correlations that bring some 
nucleons to one side and some others to another side of nucleus. 
In addition, in $N=Z$ nuclei, the neutron and proton wave functions differ only slightly because of 
the Coulomb force, which is also an implicit but dynamical correlation through the mean fields
due to the attraction between neutrons and protons. 
The deformation and its coherence between neutrons and protons together  
could lead to cluster correlations in some deformed nuclei, as will be seen later.
Furthermore, when the $nn$ and $pp$ pairing correlations are taken into account, one would expect that there will be an additional attractive correlation between spin-up and -down nucleons. The effect of pairing on the correlation between spin-up and -down will also be investigated.

We also remark on the symmetries imposed in the present calculations. 
Because of the axial symmetry around the $z$ axis, 
$|\Psi|^2$ is invariant under the simultaneous rotation of all 
the coordinates around the $z$ axis. 
Furthermore, since the present HF and HF + BCS wave functions are given as 
a product of neutron and proton parts, $|\Psi|^2$ is independent of the 
relative angle around $z$ axis between the neutron and proton coordinates. 
The same applies to the reflection symmetry as well; $|\Psi|^2$ is left unchanged 
under reflection of all the neutron and/or proton positions and spins.

\subsection{Ground states of s- and sd-shell $N=Z$ nuclei}\label{ssec:gs}

Now we present the results for the ground states of 
$^{8}$Be, $^{12}$C, $^{16}$O, $^{20}$Ne, $^{24}$Mg, and $^{28}$Si nuclei. 
Since neutron and proton wave functions are almost the same in $N=Z$ system 
and there is no explicit correlation between neutrons and protons, 
we shall show only the most probable configurations of neutrons. 

\begin{figure}
\includegraphics[width=\linewidth]{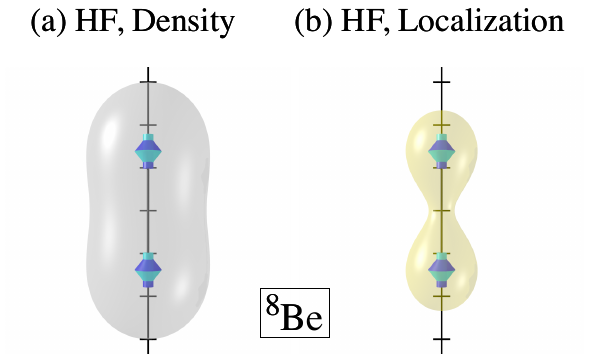}
\caption{The most probable configuration of spin-up (-down) neutrons 
in the HF ground state of $^{8}$Be are shown with blue (skyblue) arrows together with 
(a) the neutron one-body density and 
(b) the neutron localization function $\overline{\cal C}_{n\uparrow}$. 
The density isosurface in (a) is drawn at half the maximum value, 
while the localization function in (b) is drawn at 0.8 times the maximum. }
\label{fig:Be8}
\end{figure}

\begin{figure}
\includegraphics[width=\linewidth]{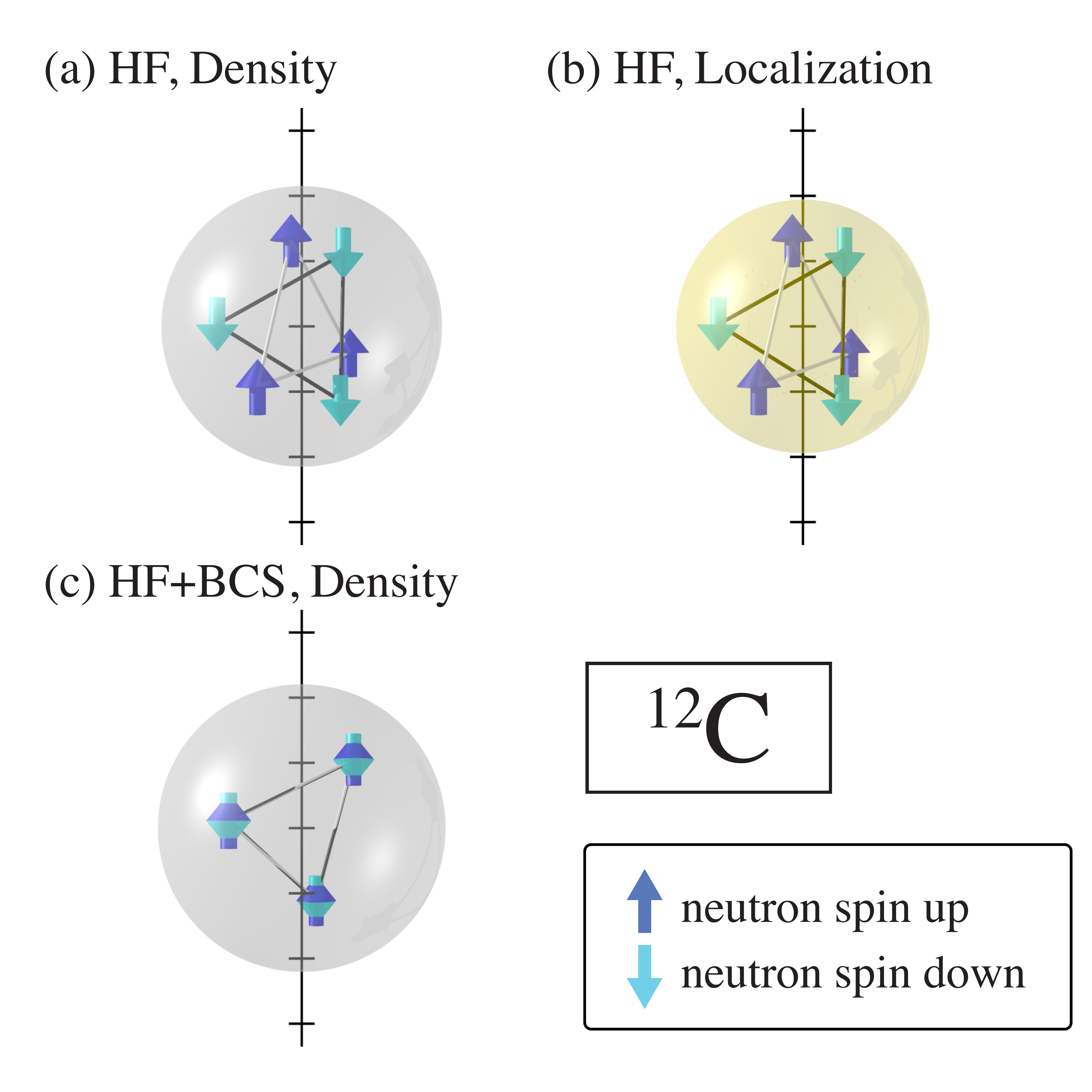}
\caption{The most probable configurations of spin-up (-down) neutrons 
obtained by the HF and the HF+BCS wave functions
in the ground state of $^{12}$C nucleus are shown with blue (skyblue) arrows.
We show
(a) the HF result together with the neutron one-body density,
(b) the HF result together with the neutron localization function, and
(c) the HF+BCS result together with the neutron one-body density.
The density isosurface is drawn at half the maximum value, 
while the localization function is drawn at 0.8 times the maximum. 
}
\label{fig:C12}
\end{figure}

\begin{figure}
\includegraphics[width=\linewidth]{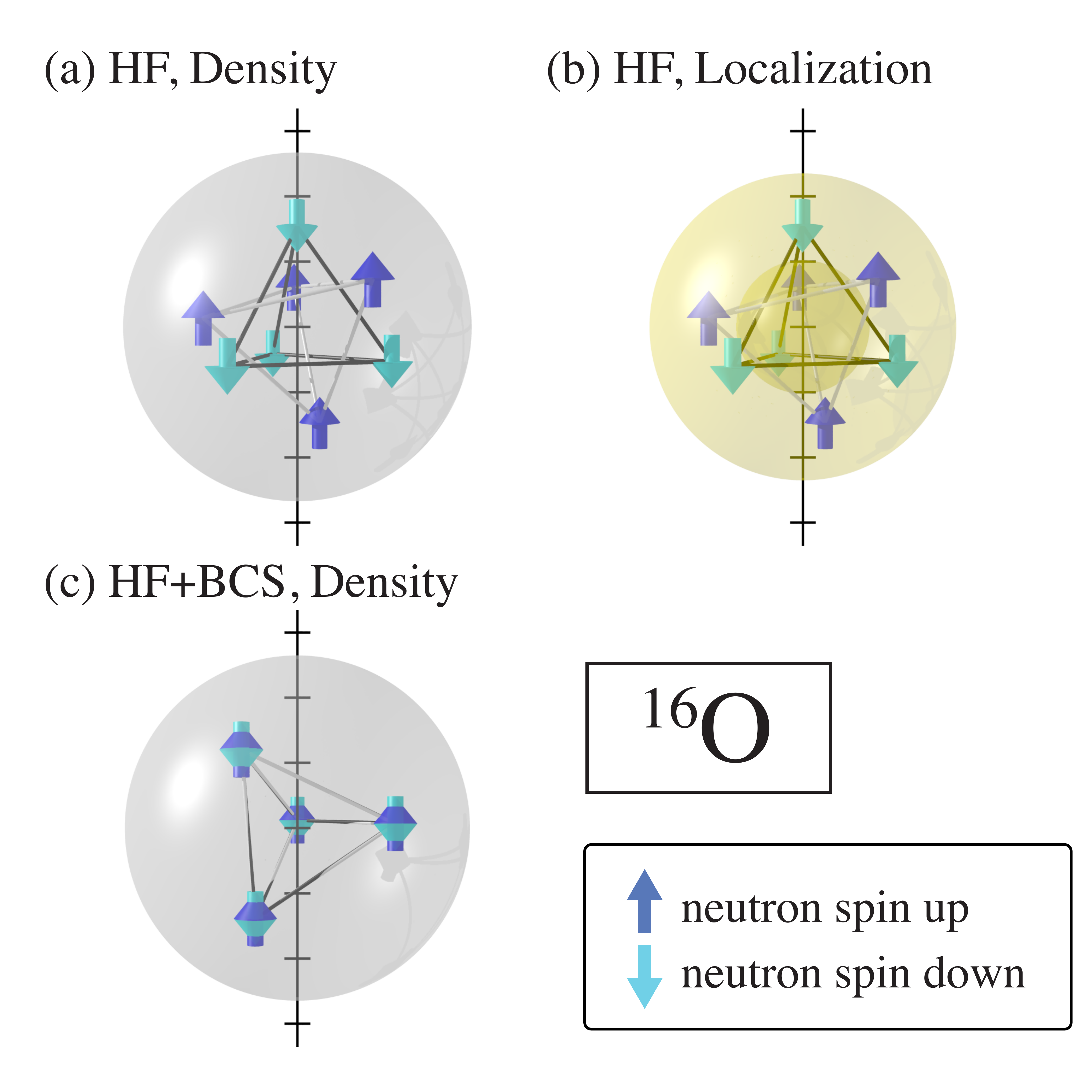}
\caption{The same as Fig. \ref{fig:C12} but for the ground state of the $^{16}$O nucleus. 
There are two surfaces for the localization function because 
it is peaked around 1.5 fm away from the center of the nucleus. }
\label{fig:O16}
\end{figure}

\begin{figure}
\includegraphics[width=\linewidth]{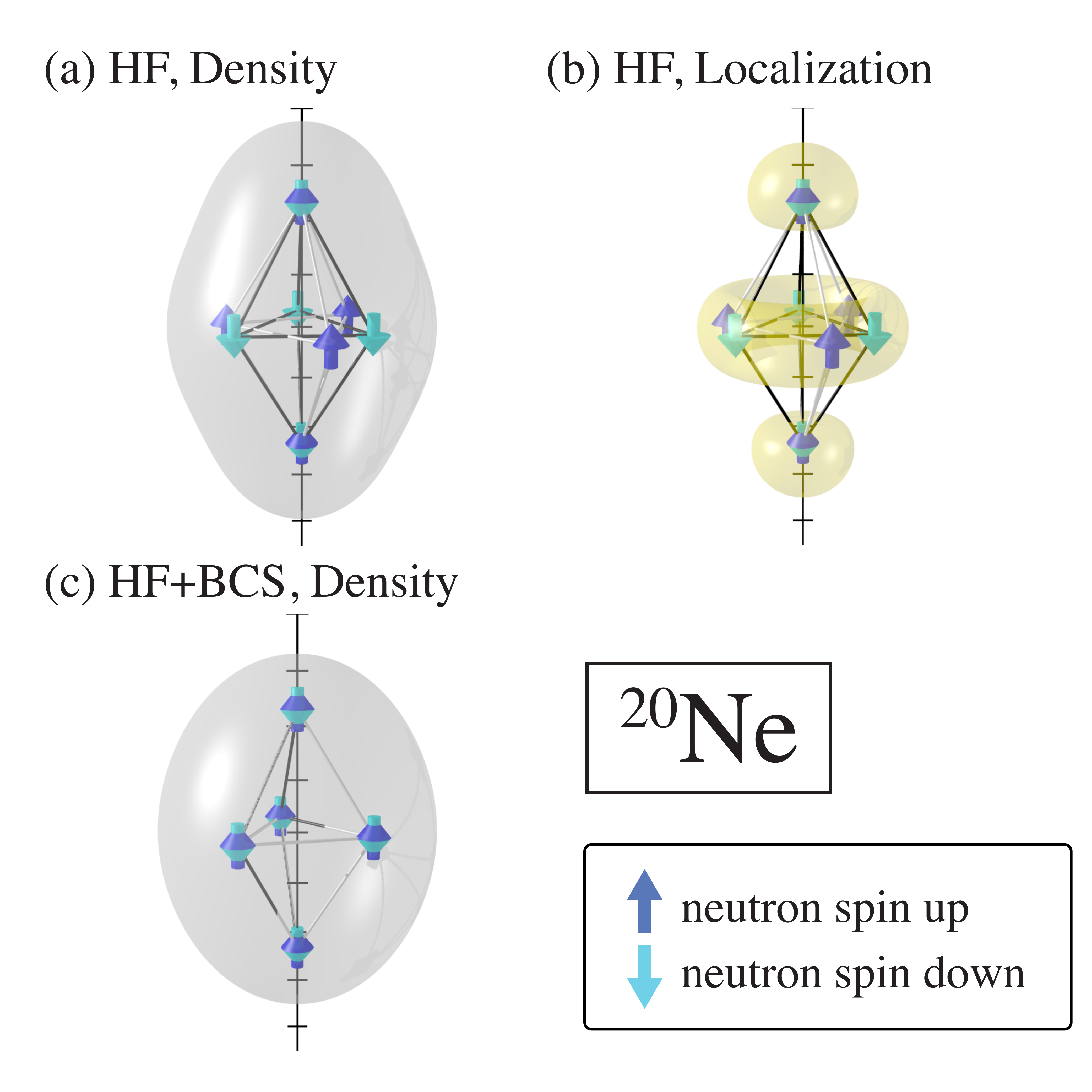}
\caption{The same as Fig. \ref{fig:C12} but for the ground state of the $^{20}$Ne nucleus.}
\label{fig:Ne20}
\end{figure}

\begin{figure}
\includegraphics[width=\linewidth]{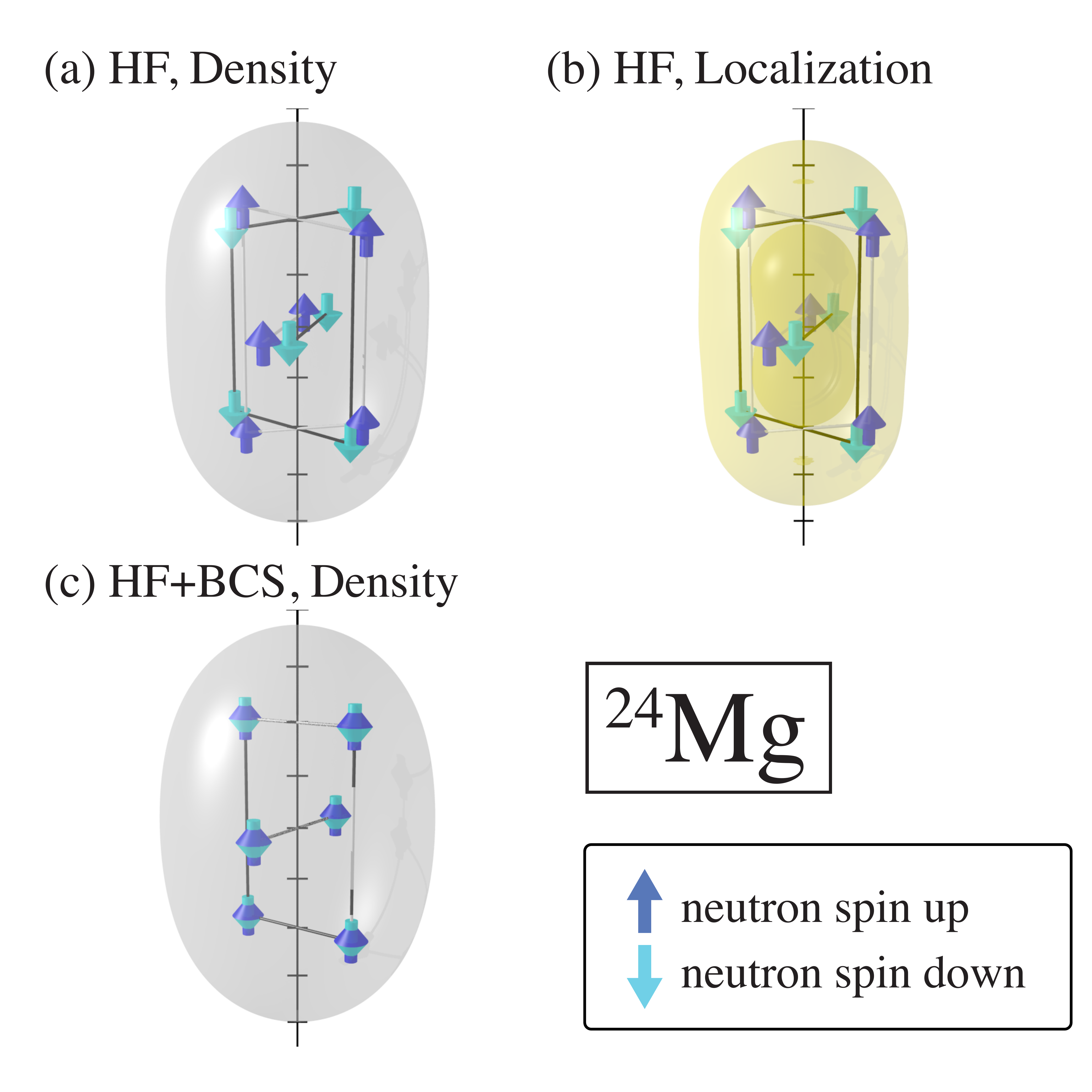}
\caption{The same as Fig. \ref{fig:C12} but for the ground state of the $^{24}$Mg nucleus.}
\label{fig:Mg24}
\end{figure}

\begin{figure}
\includegraphics[width=\linewidth]{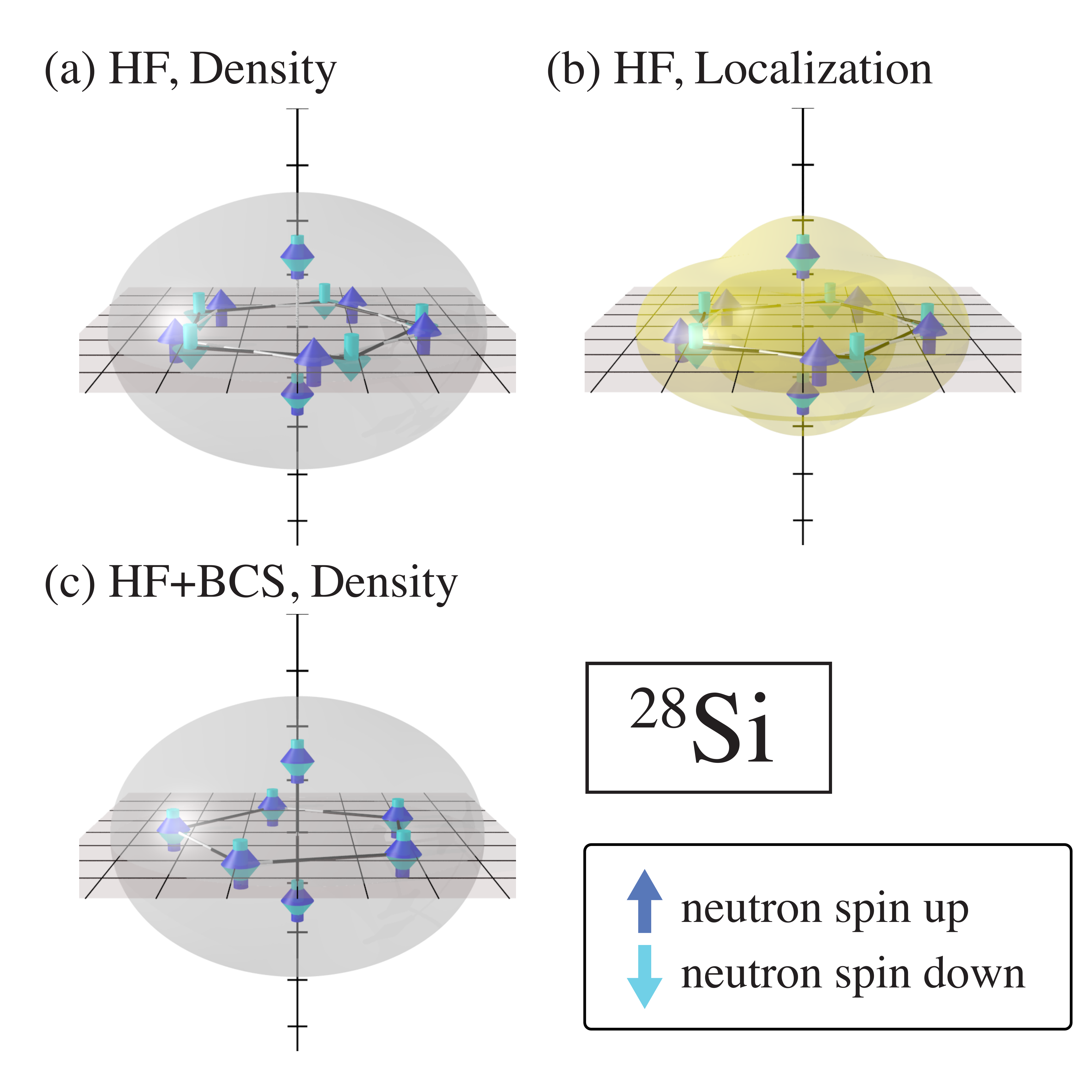}
\caption{The same as Fig. \ref{fig:C12} but for the ground state of the $^{28}$Si nucleus.}
\label{fig:Si28}
\end{figure}

Figures \ref{fig:Be8}-\ref{fig:Si28} show the most probable arrangements of 
neutrons in the mean-field ground states. The positions of spin-up (-down) neutrons 
are represented with blue (skyblue) arrows, together with the 
isosurfaces of the neutron one-body density and localization function \cite{BE90,RMUO11}. 
The density isosurface is drawn at half the maximum value while the localization 
function (multiplied by the one-body density), as defined below in Eq. \eqref{eq:Cbar}, is drawn at 0.8 times the maximum. 
The bars in the figures connecting the arrows are added to clarify the structures 
but do not have a physical significance. 
The vertical axis is the symmetry axis, and the ticks on it are located at every 1 fm. 

The localization function shown in Figs. \ref{fig:Be8}-\ref{fig:Si28}(b) is a 
measure of localization in the HF wave function, which is related to the spatial two-body correlation 
between two like-spin fermions of the same kind \cite{BE90,RMUO11,ZSN16,SN17,EKN17,JSSN18,Ta19,Ren22}. 
It was developed in quantum chemistry \cite{BE90} and introduced to nuclear physics 
in Ref. \cite{RMUO11}. 
It was recently used also to study the extra kinetic energy 
due to the Pauli exclusion during the collision process \cite{USG21}.
The localization function for a particle of kind $q = n$ 
or $p$ with spin $\sigma=\uparrow$ or $\downarrow$ is defined as
\begin{align}
{\cal C}_{q\sigma}(\bm r) = 
\left[1+\left(\frac{\rho_{q\sigma}\tau_{q\sigma}-\bm j_{q\sigma}
-\frac{1}{4}\bm\nabla(\rho_{q\sigma})^2}
{\rho_{q\sigma}\tau_{q\sigma}^{\rm TF}}\right)^2\right]^{-1}, 
\end{align}
where $\rho_{q\sigma}$, $\tau_{q\sigma}$, and $\bm j_{q\sigma}$ are 
the one-body density, kinetic-energy density, and current density of 
particle $q\sigma$, respectively, 
and $\tau_{q\sigma}^{\rm TF}=\frac{3}{5}(6\pi)^{2/3}\rho_{q\sigma}^{5/3}$ is 
the Thomas-Fermi kinetic-energy density. 
A value of ${\cal C}_{q\sigma}$ close to $1$ implies the signature of localization, 
which means that the probability of finding two particles of $q\sigma$
close to each other is very low. ${\cal C}_{q\sigma}\approx 1$ 
simultaneously for all the spin-isospin combinations is a 
minimal necessary condition of $\alpha$ clusterization \cite{RMUO11}.
In the present case with $N=Z$ and time-reversal symmetry, the wave functions of 
neutron and proton are approximately the same, and ${\cal C}_{q\uparrow}$ and 
${\cal C}_{q\downarrow}$ are exactly the same, so it suffices to consider only ${\cal C}_{n\uparrow}$. 
Since the localization is not a meaningful quantity in the regions where the one-body density is close to zero, 
we look at the localization function multiplied by the normalized one-body density, 
\begin{equation}
\overline{\cal C}_{q\sigma}(\bm r) = {\cal C}_{q\sigma}(\bm r)
\frac{\rho_{q\sigma}(\bm r)}{{\rm max}\rho_{q\sigma}(\bm r)}, 
\label{eq:Cbar}
\end{equation}
as was done in Ref. \cite{ZSN16}.

The localization function is shown for comparison with the result of $|\Psi|^2$ maximization. 
Notice that the localization function only represents the two-body correlation between 
nucleons with the same spin and tells nothing about the correlation between different spins. 
On the other hand, one can see the correlations between spin-up and -down neutrons as well 
with the $|\Psi|^2$-maximization method since it is based on the full information the wave function and takes 
into account the correlations among all nucleons. 

Figure \ref{fig:Be8} shows the result for the HF ground state of $^{8}$Be nucleus.
It is well established that the nucleus has very pronounced 2$\alpha$ structure, and its mean-field wave function indeed exhibits such structure. 
One sees that pairs of spin-up and -down 
neutrons are located at the upper and lower part of the system. 
Since the proton configuration is almost identical to that of neutrons, 
there are quartets of nucleons consisting of different spins and isospins at 
almost the same position. 
It indicates that there are two $\alpha$-cluster-like objects in this system. 
The neutron localization function, which may be regarded as a measure of $\alpha$ clustering, 
shown in Fig. \ref{fig:Be8} (b) has large values around the most probable positions 
of the neutrons. Thus one finds that the implication about $\alpha$ clustering of the 
localization function is consistent with the picture obtained with the present approach 
in this particular case. 

Figure \ref{fig:C12} shows the result for the $^{12}$C nucleus.
The $^{12}$C nucleus is spherical in the present calculations. 
As seen in Figs. \ref{fig:C12}(a) and (b) for the HF case, 
the most probable positions of the three neutrons with the same spin form an equilateral triangle. 
We have found that, in the HF case, there is almost no correlation in  
the relative angle about $z$ axis between the two triangles. 
With the pairing correlation, the two triangles are overlaid in the 
most probable arrangement, as seen in Fig. \ref{fig:C12}(c) for the HF+BCS case. 
Thus it turns out, as could be expected, that the pairing correlation induces an 
attractive correlation between spin-up and -down nucleons. 
We recall that the protons are not explicitly correlated with the neutrons. 
Thus, in the present HF and HF + BCS wave functions there are no $\alpha$-like 
correlation as observed in the $^8$Be case. 
In the present calculation, only if there is a pair of neutrons on the symmetry 
axis and $N=Z$ can one say that there is an $\alpha$-cluster-like four-body correlation. 

In $^{16}$O nucleus shown in Fig. \ref{fig:O16}, 
the four neutrons with the same spin form a regular tetrahedron 
[Figs. \ref{fig:O16}(a) and (b)]. 
As in the case of $^{12}$C nucleus, only with the pairing correlation do
the spin-up and -down tetrahedrons favor the same orientation. 
Again, since the protons are not correlated with the neutrons, one cannot 
claim $\alpha$-cluster correlation in our present framework. 
Note that we forced the pairing gaps to be nonzero by the constant-gap 
approximation although the pairing is not likely to be active in this nucleus. 

\begin{figure}
\includegraphics[width=\linewidth]{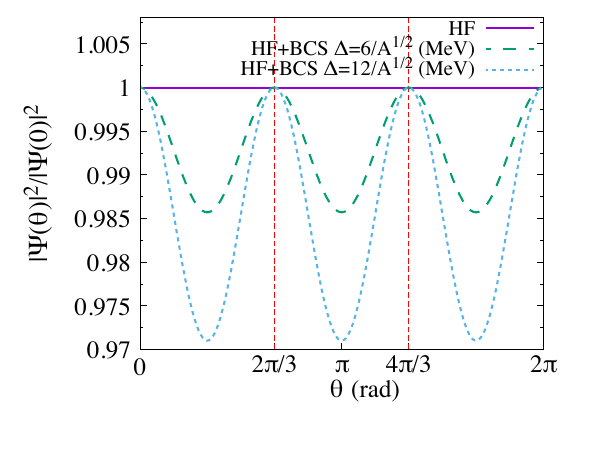}
\caption{
$|\Psi|^2$ of $^{20}$Ne as functions of the relative angle $\theta$ between 
spin-up and -down neutrons. The angle $\theta$ is defined as the rotation angle 
of the positions of spin-up neutrons about the $z$ axis while those of spin-down neutrons are fixed. $\theta = 0$ corresponds to the most probable configuration shown in Fig. \ref{fig:Ne20}. 
The probabilities are normalized with their own maximum values. 
The solid curve represents the HF result, 
and the dashed and dotted curves show the HF+BCS results with 
pairing gaps $\Delta=6/A^{1/2}$ MeV and $\Delta=12/A^{1/2}$ MeV, respectively.}
\label{fig:Ne20_rotate}
\end{figure}

Figure \ref{fig:Ne20} shows the result for the $^{20}$Ne nucleus. 
In contrast to the two spherical nuclei $^{12}$C and $^{16}$O, 
$^{20}$Ne is deformed, and interesting cluster structure associated with it is observed with our method. 
As seen in Fig. \ref{fig:Ne20}, there are two $\alpha$ clusters at the top and bottom along the $z$ axis, 
and at the middle are two regular triangles of spin-up and -down neutrons, respectively. 
As can be seen in Fig. \ref{fig:Ne20}(b), the nucleons are most likely to be
located at the regions where the localization function has large values. 
However, despite the large values of the localization function at the middle 
of the system, we conclude from our analysis that $\alpha$ particles are not 
likely to exist around there in the present mean-field wave functions. 
As we shall see later, without pairing, there is no rotational correlation around the $z$ axis between the spin-up and -down triangles. 
Moreover, by construction there is no such correlation between neutron and proton triangles as mentioned in the previous subsection. 
We can only state that in the present setup that each triplet of nucleons 
with the same spin is likely to form a regular triangle and that 12 nucleons 
are likely to be located along the circle 
on the $xy$ plane. With pairing, spin-up and -down triangles of the same 
nucleon kind favor the same orientation (see Fig. \ref{fig:Ne20_rotate}). 
If we took into account explicit neutron-proton correlations via, {\it e.g.}, 
neutron-proton pairing or HF quasiparticles with isospin mixing 
\cite{Pe04,Sa13,Sh14,FM14} in addition to the pairing between like nucleons, 
there might be $\alpha$ clusters in the middle region as well as the top and bottom. 
We also note that the ``bipyramidal'' structure seen in our results supports the assumption of 
the cluster model employed in Ref. \cite{BI21}, if one is concerned only with 
neutrons or protons.

Figure \ref{fig:Mg24} shows the results for the $^{24}$Mg nucleus. 
In the HF case [Fig. \ref{fig:Mg24}(a)], four of the spin-up neutrons form a trapezoid, 
while the remaining two are located at both sides of it. 
The spin-down neutrons are arranged symmetrically with the spin-up ones. 
As a result, there are six close pairs of neutrons. 
As seen from the Fig. \ref{fig:Mg24}(b) the neutrons are located in the region where 
the localization function is larger. 
With pairing [Fig. \ref{fig:Mg24}(c)], the two trapezoids become rectangles, and the arrangements of spin-up and -down nucleons become identical.

Figure \ref{fig:Si28} shows the results for the $^{28}$Si nucleus, which is oblately deformed.
It is likely that there are two $\alpha$ clusters along the $z$ axis, and 
regular pentagons are formed on the $xy$ plane by the remaining nucleons with the same spin. 
The two pentagons coincide with each other with pairing correlations. 

\begin{figure}
    \includegraphics[width=\linewidth]{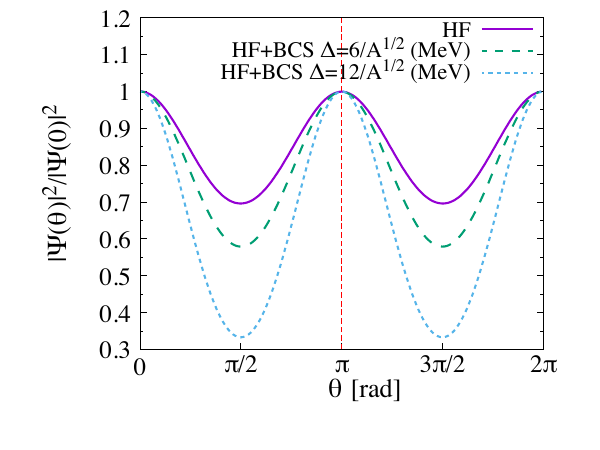}
    \caption{The same as Fig. \ref{fig:Ne20_rotate} but for the ground state of the $^{24}$Mg nucleus.}
    \label{fig:Mg24_rotate}
\end{figure}

Next, we investigate the effects of the pairing correlation 
by observing the behavior of the probability distribution around the maximum. 
We pick up $^{20}$Ne and $^{24}$Mg, and we rotate only the positions of spin-up neutrons by angle $\theta$ about the $z$ axis while spin-down neutrons are fixed. 
Figures \ref{fig:Ne20_rotate} and \ref{fig:Mg24_rotate} show comparisons of 
the $\theta$ dependence of $|\Psi|^2$ with and without the pairing in $^{20}$Ne and $^{24}$Mg, respectively. 
The solid curve shows the probability relative to the maximum in the HF case, while 
the dashed and dotted curves show the same thing with different values of the pairing gap, 
$\Delta=6/A^{1/2}$ MeV and $\Delta=12/A^{1/2}$ MeV, respectively. 
Since the most probable neutron configuration of the $^{20}$Ne ($^{24}$Mg) nucleus is 
symmetric under rotation by $2\pi/3$ ($\pi$), $|\Psi|^2$ changes periodically. 
As can be seen in the two figures, as the pairing strength increases, the probability variation 
also increases. 
We have also tried a similar analysis on the spherical $^{12}$C nucleus by rotating the spin-up neutrons around the $z$ axis while the spin-down neutrons are fixed. We have observed, without pairing, that the probability is nearly constant whereas with pairing the probability changes by $\approx  60$ \%. 

To briefly summarize the above results, we saw the $\alpha$-like four-body correlations, which are induced by collective deformations and its coherence between neutrons and protons, along the $z$ axis in 
some of the deformed nuclei. It is also found that the pairing correlation always reinforces the attractive correlations between spin-up and -down nucleons. 
It is remarkable that the correlation induced by the deformation is 
reflected in the most likely configuration of nucleons as the cluster-like correlation.

\subsection{Deformed states of $^{16}$O}\label{ssec:O16}

\begin{figure}
\includegraphics[width=\linewidth]{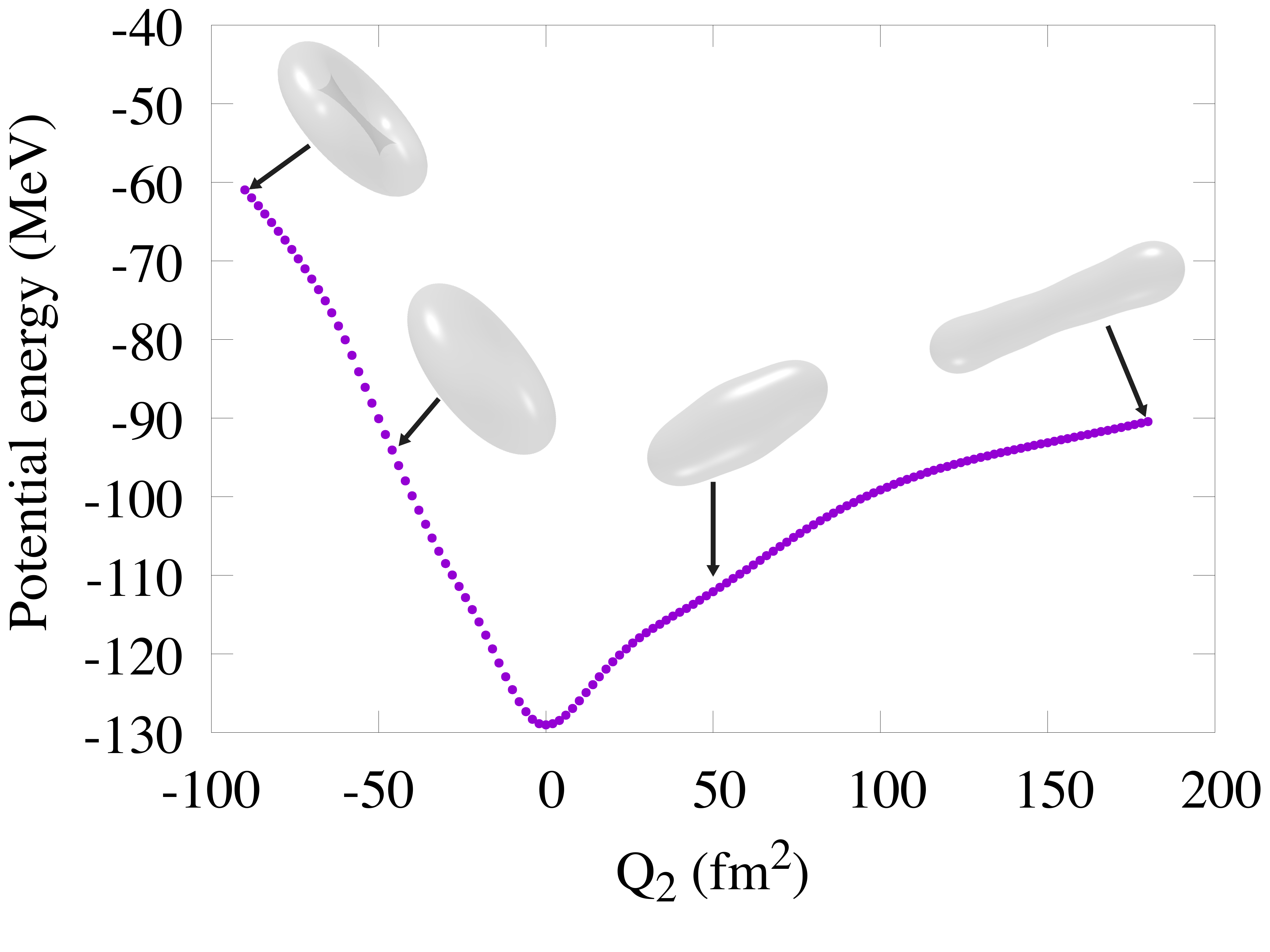}
\caption{The HF+BCS potential-energy curve of the $^{16}$O nucleus as a 
function of the quadrupole moment $Q_2$. 
The HF + BCS result is obtained with the constant-gap approximation 
with the gap parameter $\Delta=12/A^{1/2}$ MeV. 
The density isosurfaces at $Q_2=-90$, $-50$, $50$, and $180$ fm$^2$ are also shown.}
\label{fig:pes_O16}
\end{figure}

Here we investigate the deformed states of $^{16}$O nucleus 
obtained with constrained HF + BCS calculations. 
In Fig. \ref{fig:pes_O16} we show the potential-energy curve of the nucleus 
as a function of the quadrupole deformation $Q_2$ defined as 
\begin{eqnarray}
Q_2 = \int d^3r\ r^2Y_{20}(\hat{\bm r})\rho(\bm r), 
\end{eqnarray} 
where $\rho(\bm r)$ is the total one-body density. 
The ground state is spherical, and as the deformation increases in the prolate direction, 
there appears an elongated shape with a bulge at the middle and eventually a rod shape, 
while at large oblate deformations there appears a torus shape. 

\begin{figure}
\includegraphics[width=\linewidth]{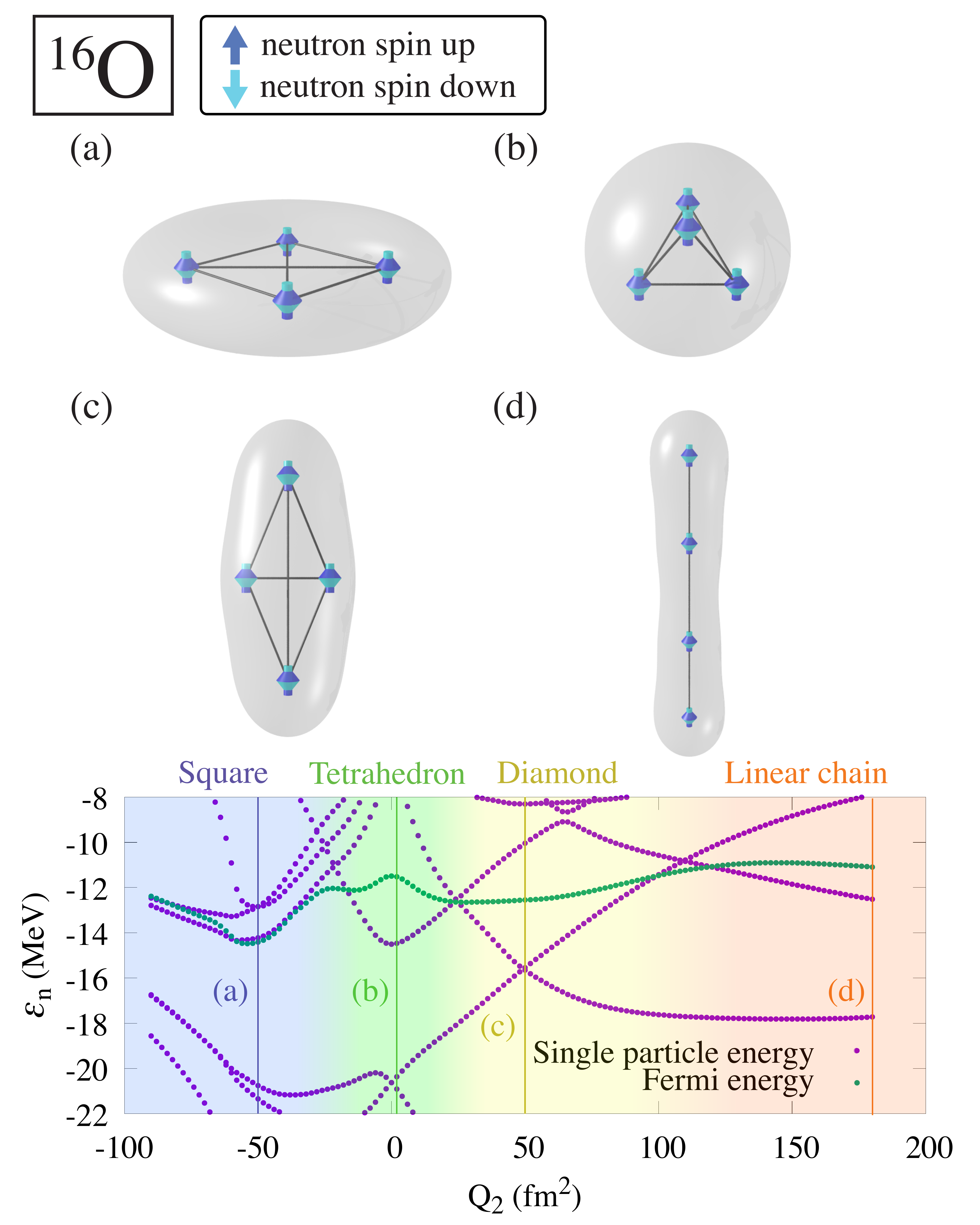}
\caption{The most probable neutron configurations of the $^{16}$O nucleus at (a) $Q_2 = 50$ fm$^2$, 
(b) 2 fm$^2$, (c) 50 fm$^2$, and (d) 180 fm$^2$. The lower panel shows the 
neutron single-particle energies and the Fermi energy as functions of $Q_2$. 
The background colors are inserted according to the four regimes of neutron configurations; 
square for $Q_2 \lesssim -25$ fm$^2$, tetrahedron for 
$-25\ {\rm fm}^2 \lesssim Q_2 \lesssim 25\ {\rm fm}^2$, 
diamond for $25\ {\rm fm}^2 \lesssim Q_2\lesssim 100\ {\rm fm}^2 $, and 
linear chain for $100\ {\rm fm}^2 \lesssim Q_2$. 
The vertical lines labeled by (a)-(d) correspond to the $Q_2$ values at which the 
upper figures are obtained. 
}
\label{fig:O16_def}
\end{figure}

We performed the $|\Psi|^2$ maximization for the $^{16}$O nucleus with different $Q_2$ values. 
The resulting most probable neutron configurations for four selected values 
of $Q_2$ are shown in Fig. \ref{fig:O16_def} 
together with the neutron single-particle energies near the Fermi energies 
as functions of $Q_2$. 
It is found that there are four regimes of the nucleon arrangement depending on the occupied single-particle levels. 
We have observed abrupt transitions of the neutron arrangement around the 
$Q_2$ values where an alternation of the Fermi level occurs: 
$Q_2\approx -25$ fm$^2$, $25$ fm$^2$, and $100$ fm$^2$. 
In the most oblate region where $Q_2\lesssim -25$ fm$^2$, the four pairs of 
neutrons are most likely to form a square on the $xy$ plane [Fig. \ref{fig:O16_def}(a)]. 
In the near-spherical region, the pairs form a tetrahedron [Fig. \ref{fig:O16_def}(b)]. 
In the prolate region where $25\lesssim Q_2\lesssim 100$ fm$^2$, 
the pairs form a diamond shape aligned to $z$ axis, indicating there are 
$\alpha$ clusters at the top and the bottom [Fig. \ref{fig:O16_def}(c)]. 
Finally, in the most prolate region ($Q_2\gtrsim 100$ fm$^2$), all the pairs are located on 
the $z$ axis, forming a four-$\alpha$ linear chain. 

From the various structures obtained by the $|\Psi|^2$ maximization, 
we found the characteristic arrangements of nucleons in different regimes of 
deformation or occupied orbitals, which was not clear if we merely observed the shapes. 

\section{Summary and future perspectives}\label{sec:summary}

In this paper, we have introduced a method to visualize the many-body correlations 
based on the full information of the wave function, and demonstrated the usefulness of the 
method in nuclear physics. 
The method visualizes the set of nucleon coordinates which maximizes the square of the 
many-body wave function. 
We have applied it to HF and HF + BCS wave functions of $p$- and $sd$-shell $N=Z$ even-even nuclei 
to study the cluster and other correlations in these systems. 
It was found that the HF wave function already contains 
$\alpha$-cluster-like correlations in some deformed nuclei. 
From the HF + BCS wave functions we found that 
the pairing correlations induce attractive correlations between spin-up 
and -down nucleons. 

We believe that the present method gives a new viewpoint to 
the microscopic nuclear wave function. 
There are several directions for further developments of such methods. 
\begin{enumerate}
\item Analyses of more global behaviors of the wave function

In the present work, we only searched for the maximum of $|\Psi|^2$, 
but it would be necessary and interesting to study more global behaviors 
as was mentioned in Sec. \ref{sec:method}. 
The Markov-chain Monte-Carlo method would be useful to analyze
complicated structures, such as fluctuations and local maxima, 
of the probability distribution in the multi-dimensional space. 

\item Applications to more correlated many-body states 

The merit of our method is fully exploited when it is applied to a ``black box'' 
wave function that contains rich and nontrivial correlations. 
Thus the shell model or {\it ab initio} wave functions, which take into account 
correlations indiscriminately, are more suited for the method.
It would also be interesting to analyze the structure of 
the random-phase approximation or generator-coordinate method wave functions 
that take into account the correlations associated with collective motions \cite{RS,Rowe}. 
The collective motion would induce fluctuations of the collective degrees 
of freedom including the cluster structure. 

\item Studies of phenomena other than clustering

There are several phenomena that can be analyzed with the present method or its extensions. 
The molecular-bond structure accompanying clusters and the correlations among 
valence neutrons in neutron-rich nuclei can be interesting targets of our analyses. 
Visualization of motions of individual nucleons during reaction dynamics would also 
help understanding of mechanisms of fusion, fission, and multi-nucleon transfer reactions. 

\end{enumerate}

\acknowledgments
The authors are thankful for discussions with Yu Liu, one of the authors of 
Refs. \cite{Liu16,Liu19,Liu20_JPCL,Liu20_NC}. 
This work was supported by the JSPS KAKENHI  Grant No. 19K03861. 
M. M. acknowledges support from Graduate Program on Physics for the Universe (GP-PU) 
of Tohoku University.

\appendix

\section{$d_s$ dependence of $\rho^{(N)}_{d_s}$}\label{app:ds}

In this appendix, we demonstrate numerically the validity of the assumption made 
in Sec. \ref{sec:results} that the maximum of $\rho^{(N)}_{d_s=0}(\bm r_1,\dots,\bm r_N)$ 
[Eq. \eqref{eq:rhoN_ds}] is the global maximum of $\rho^{(N)}(x_1,\dots,x_N)$ [Eq. \eqref{eq:rhoN}], taking $^{20}$Ne as an example. 

\begin{figure}
\includegraphics[width=\linewidth]{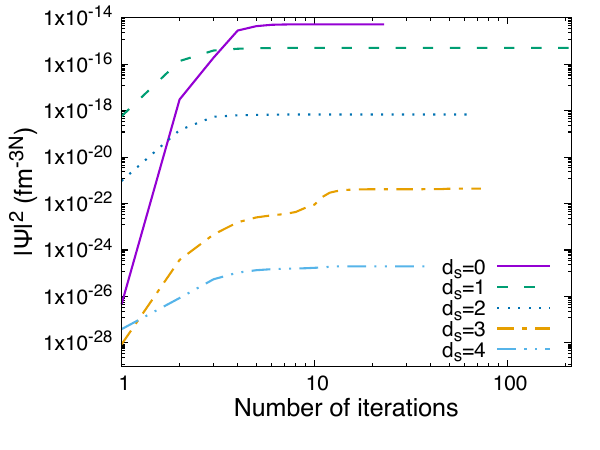}
\caption{Values of $\rho^{(N)}_{d_s}$ for $d_s=0,\dots,4$ as functions of CG iterations for the neutron part of the wave function of the $^{20}$Ne nucleus. The initial spatial neutron coordinates are generated randomly. }
\label{fig:ds}
\end{figure}

In Fig. \ref{fig:ds}, we show the values of $\rho^{(N)}_{d_s}$ with $0\leq d_s\leq 4$ 
as functions of CG iterations for the neutron part of the HF wave function of 
$^{20}$Ne nucleus. 
Although the initial values and the convergence behaviors are different, 
$\rho^{(N)}_{d_s=0}$ converges indeed to the largest value, and the others converge to 
values smaller by orders of magnitude. 
It can also be seen that the maximum values of $\rho^{(N)}_{d_s}$ decreases with 
the absolute value of $d_s$. 

This example shows the validity of the assumption; 
it is most likely that the numbers of spin-up and -down nucleons are equal in the 
ground state so that the nucleons gain energy from the short-range and attractive 
nucleon-nucleon interaction.

\section{BCS wave function in coordinate-space representation}\label{app:bcs}

In this appendix, we give the expression of the $N$-particle component of a 
BCS wave function in the coordinate-space representation, written in terms of 
the Pfaffian of a certain matrix.
Although the coordinate-space BCS or so-called APG (antisymmetrized product of geminals)
wave functions have been discussed and utilized in
the context of condensed-matter physics and chemistry \cite{BGL88,Ba06,Ba08,Geno20}, 
the notations and structures of their wave functions are suited to many-electron systems rather than nuclear systems. 
Here we present similar formulas that conform to the nuclear-physics notation. 
It is remarkable that the formula [Eq. \eqref{eq:bcsxfin}], given in the form of a Pfaffian, 
represents the natural generalization of the Slater determinant for HF 
to the generalized product wave function for BCS or Hartree-Fock-Bogoliubov (HFB) theory \cite{RS}, including the HF Slater determinant as a special case. 
We start with a simpler case where all the fermions are paired, 
then we handle a more complicated situation where there are unpaired particles 
as well as pairs. 

\subsection{Fully paired state}
We consider the usual BCS state expressed as 
\begin{eqnarray}
\left|\Psi\right\rangle&=&\prod_{k>0}(u_k+v_k a^{\dagger}_k a^\dagger_{\bar{k}})\left|0\right\rangle
\nonumber\\
&\propto&
\exp\left(\sum_{k>0}\frac{v_k}{u_k}a^{\dagger}_k a^\dagger_{\bar{k}}\right)|0\rangle, 
\label{eq:bcs0}
\end{eqnarray}
where $k>0$ labels a single-particle state, $\bar k\equiv-k$ denotes the ``conjugate'' state to be paired with $k$ \cite{RS}, and $|0\rangle$ is the bare vacuum. 
The coefficients $u_k$ and $v_k$ are, in general, complex variational parameters. 
The exponent of Eq. \eqref{eq:bcs0} can be rewritten as 
\begin{eqnarray}
&&
\sum_{k>0}\frac{v_k}{u_k}a^{\dagger}_k a^\dagger_{\bar{k}}
\nonumber\\
&=&
\int dxdx'\ \sum_{k>0}\frac{v_k}{u_k}\varphi_k(x)\varphi_{\bar k}(x')
\psi^{\dagger}(x)\psi^{\dagger}(x')
\nonumber\\
&=&
\int dxdx'\ Z(x,x')\psi^{\dagger}(x)\psi^{\dagger}(x'), 
\end{eqnarray}
where $\varphi_k(x)$ is the single-particle wave function, $\psi(x)$ is the nucleon 
field operator, and 
\begin{eqnarray}
Z(x,x')\equiv\sum_{k>0}\frac{v_k}{u_k}\varphi_k(x)\varphi_{\bar k}(x'). 
\label{eq:Z}
\end{eqnarray}
Note that $x$ denotes the position and the spin variables. 
\begin{widetext}
The $N$-particle component $|\Psi_N\rangle$, where $N$ is even, of the BCS state as given in 
Eq. \eqref{eq:bcs0} is given up to an unimportant factor by 
\begin{eqnarray}
|\Psi_N\rangle 
&\propto&
\left(\sum_{k>0}\frac{v_k}{u_k}a^{\dagger}_k a^\dagger_{\bar{k}}\right)^{N/2}|0\rangle
\nonumber\\
&=&
\left(\int dxdx'\ Z(x,x')\psi^{\dagger}(x)\psi^{\dagger}(x')\right)^{N/2}|0\rangle
\nonumber\\
&=&
\int dx_1dx_2\dots dx_{N-1}dx_N\ 
Z(x_1,x_2)\dots Z(x_{N-1},x_N)
\psi^{\dagger}(x_1)\psi^{\dagger}(x_2)\dots
\psi^{\dagger}(x_{N-1})\psi^{\dagger}(x_N)|0\rangle. 
\end{eqnarray}
It can then be shown by using the anticommutation relation of the fermion operators 
that the coordinate-space wave function is given by 
\begin{eqnarray}
\langle x_1,\dots,x_N|\Psi\rangle &=&
\langle 0|\psi(x_1)\dots\psi(x_N)|\Psi_N\rangle
\nonumber\\
&\propto&
\sum_{\sigma\in S_N}{\rm sgn}(\sigma)
Z(x_{\sigma(1)},x_{\sigma(2)})Z(x_{\sigma(3)},x_{\sigma(4)})
\dots Z(x_{\sigma(N-1)},x_{\sigma(N)}), 
\end{eqnarray}
where $S_N$ is the symmetric group on $N$ objects, 
and ${\rm sgn}(\sigma)$ equals $+1$ ($-1$) for even (odd) permutations. 
The above expression can be rewritten as 
\begin{eqnarray}
\langle x_1,\dots,x_N|\Psi\rangle
&\propto&
\frac{1}{(N/2)!}
\sum_{\sigma\in S_N'}{\rm sgn}(\sigma)
\bar Z(x_{\sigma(1)},x_{\sigma(2)})\bar Z(x_{\sigma(3)},x_{\sigma(4)})
\dots \bar Z(x_{\sigma(N-1)},x_{\sigma(N)}), 
\label{eq:bcs0x}
\end{eqnarray}
where 
\begin{eqnarray}
\bar Z(x,y) \equiv Z(x,y) - Z(y,x), 
\end{eqnarray}
and 
\begin{eqnarray}
S_{2n}' = \left\{\sigma\in S_{2n}|\sigma(2i-1)<\sigma(2i)\ (1\leq i\leq n)\right\}. 
\end{eqnarray}
%
%
The right hand side of Eq. \eqref{eq:bcs0x} is nothing but the Pfaffian of the 
following skew symmetric $N\times N$ matrix \cite{IW95}, 
\begin{eqnarray}
\mathcal{Z}=\left(
\begin{array}{cccc}
0 & \bar{Z}_{12} & \ldots&\bar{Z}_{1N} \\
\bar{Z}_{21}&0&\ldots&\bar{Z}_{2N} \\
\vdots&\vdots&\ddots&\vdots \\
\bar{Z}_{N1}& \bar{Z}_{N2}&\ldots&0 \\
\end{array}
\right),
\end{eqnarray}
where $\bar Z_{ij} \equiv \bar Z(x_i,x_j) = -\bar Z_{ji}$. 
Therefore, the BCS wave function is given by a Pfaffian, 
\begin{eqnarray}
\Psi(x_1,\dots,x_N)\propto {\rm pf}{\cal Z}. 
\label{eq:bcs0pf}
\end{eqnarray}
Due to the property of a Pfaffian that $({\rm pf} A)^2 = \det A$, 
\begin{eqnarray}
\rho^{(N)}(x_1,\dots,x_N) &=& |\Psi(x_1,\dots,x_N)|^2
\propto |{\rm pf}{\cal Z}|^2 = |\det{\cal Z}|. 
\end{eqnarray}

\subsection{State with unpaired particles}

In practical calculations for nuclear systems, the levels far below the fermi level are 
unpaired and fully occupied, and the $v/u$ factors for these levels tend to diverge. 
Moreover, for odd nuclei, the last nucleon remains unpaired 
and blocks a level from the pairing correlation. 
In such cases, Eq. \eqref{eq:bcs0pf} is not applicable anymore.
Thus now we consider a more general case where $M<N$ single-particles states 
are unpaired and fully occupied. Such a state is given by 
\begin{equation}
\left|\Psi\right\rangle=\prod_{i=1}^M a^{\dagger}_i \prod_{j\geq M+1}(u_j+v_j a^{\dagger}_j a^\dagger_{\bar{j}})\left|0\right\rangle. 
\label{eq:bcs}
\end{equation}
If $M$ is an odd number, $N$ is also odd, and $N-M$ is still even. 
In the same way as the preceding discussion, 
one obtains for the coordinate-space representation of the state 
as given in Eq \eqref{eq:bcs}, 
\begin{eqnarray}
\langle x_1,\dots,x_N|\Psi\rangle
&\propto&
\sum_{\sigma\in S_N}{\rm sgn}(\sigma)
\varphi_1(x_{\sigma(1)})\dots\varphi_M(x_{\sigma(M)})
Z(x_{\sigma(M+1)},x_{\sigma(M+2)})\dots Z(x_{\sigma(N-1)},x_{\sigma(N)}). 
\label{eq:bcsx}
\end{eqnarray}
Note that the summation in the function $Z$ [Eq. \eqref{eq:Z}] in this case 
is taken over the levels other than the $M$ fully occupied or blocked ones. 
Since the number of terms in the summation grows factorially with the number of particles, 
it is difficult to make a brute-force computation for many-particle systems. 
We shall show in the following that the right hand side of Eq. \eqref{eq:bcsx} is proportional 
to a single Pfaffian of an $(N+M)\times(N+M)$ matrix, which is determined by the $\varphi$'s and $Z$'s. 
Thus $|\Psi|^2$ is easily obtained by computing a determinant. 

We will rewrite the right-hand side of Eq. \eqref{eq:bcsx} as a sum of products of a determinant and a Pfaffian. 
To this end, we first introduce a notation to specify a submatrix of a matrix. 
Consider an $m\times n$ matrix $X=(x_{ij})_{1\leq i \leq m,1\leq j \leq n}$. 
Given a subset of $r~(\leq m)$ row indices $I=\{i_1,i_2,\dots,i_r\}$ and a subset of 
$s~(\leq n)$ column indices $J=\{j_1,j_2,\dots,j_s\}$, we denote the $r\times s$ 
submatrix of $X$ by
\begin{eqnarray}
X(I;J) = (x_{i_pj_q})_{1\leq p \leq r,1\leq q \leq s}. 
\end{eqnarray}
Next, we define a notation for a subsequence of a sequence of integers. 
Let $[N]$ be the sequence of positive integers up to $N$, $\{1,2,\dots,N\}$. 
The set of the subsequences of length $r$ of $[N]$ is 
denoted by $\displaystyle \left(\begin{array}{c}[N]\\r\end{array}\right)$. 
That is, $\displaystyle \left(\begin{array}{c}[N]\\r\end{array}\right)$ contains 
all the sequences $\{k_1,k_2,\dots,k_r\}$ such that $1\leq k_1<k_2<\dots\leq N$.
Thus each element $K\in\left(\begin{array}{c}[N]\\r\end{array}\right)$ specifies a certain partition of $[N]$ into $r$ numbers and $N-r$ numbers. 
The complementary sequence of $K$ in $[N]$ is denoted by $K^c$,  which is also arranged in the ascending order. 

Let $B$ be an $N\times M$ matrix defined as 
\begin{eqnarray}
B=
\left(\begin{array}{ccc}
\varphi_{1}(1) & \cdots & \varphi_{M}(1) \\
\varphi_{1}(2) & \cdots & \varphi_{M}(2) \\
\vdots &  & \vdots \\
\varphi_{1}(N) & \cdots & \varphi_{M}(N)
\end{array}\right), 
\end{eqnarray} 
where $\varphi_i(j) \equiv \varphi_i(x_j)$. 
Using the notations defined above, one has 
\begin{eqnarray}
&&
\sum_{\sigma\in S_N}{\rm sgn}(\sigma)
\varphi_1(x_{\sigma(1)})\dots\varphi_M(x_{\sigma(M)})
Z(x_{\sigma(M+1)},x_{\sigma(M+2)})\dots Z(x_{\sigma(N-1)},x_{\sigma(N)})
\nonumber\\
&=&
\sum_{K\in\left(\begin{array}{c}[N]\\N-M\end{array}\right)}{\rm sgn}(K)~
{\rm pf}{\cal Z}(K;K)~\det B(K^c;[M]), 
\label{eq:bcsx2}
\end{eqnarray}
where ${\rm sgn}(K)$ is the sign of the permutation $\{k_1,\dots,k_{N-M},k_1^c,\dots,k_M^c\}$ 
given by concatenating $K$ and $K^c$. 
See Table \ref{tb:K} for the list of all the possible sequences 
$K\in\left(\begin{array}{c}[N]\\N-M\end{array}\right)$, 
and corresponding $K^c$ and ${\rm sgn}(K)$ for a simple case where $N=4$ and $M=2$. 
In Eq. \eqref{eq:bcsx2}, ${\rm pf}{\cal Z}(K;K)$ is the Pfaffian of an $(N-M)\times(N-M)$ submatrix of ${\cal Z}$ 
while $\det B(K^c;[M])$ is the determinant of an $M\times M$ submatrix of $B$. 
Hereafter, we keep $K$ for an element of $\left(\begin{array}{c}[N]\\N-M\end{array}\right)$. 

\begin{table}
\begin{tabular}{ccc}
\hline\hline
$K$ & $K^c$ & ${\rm sgn}(K)$ \\
\hline
$(1,2)$ & $(3,4)$ & $+1$ \\
$(1,3)$ & $(2,4)$ & $-1$ \\
$(1,4)$ & $(2,3)$ & $+1$ \\
$(2,3)$ & $(1,4)$ & $+1$ \\
$(2,4)$ & $(1,3)$ & $-1$ \\
$(3,4)$ & $(1,2)$ & $+1$ \\
\hline\hline
\end{tabular}
\caption{Subsequences $K\in \left(\begin{array}{c}[N]\\N-M\end{array}\right)$, and corresponding $K^c$ and ${\rm sgn}(K)$ for $N=4$ and $M=2$. }
\label{tb:K}
\end{table}

Now we define the following $N\times(N+M)$ matrix,
\begin{eqnarray}
T= (\mathbbm{1}_N,B)
=
\left(\begin{array}{ccccccc}
1 &   &        &    & \varphi_1(1) & \dots & \varphi_M(1) \\
  & 1 &        &    & \varphi_1(2) & \dots & \varphi_M(2) \\
  &   & \ddots &    & \vdots         &       & \vdots         \\
  &   &        &  1 & \varphi_1(N) & \dots & \varphi_M(N)
\end{array}\right), 
\end{eqnarray}
where $\mathbbm{1}_N$ is the $N\times N$ identity matrix. 
The submatrix $T([N];K\cup\{N+1,\dots,N+M\})$ 
represents an $N\times N$ matrix, 
\begin{eqnarray}
T([N];K\cup\{N+1,\dots,N+M\}) = ( I'(K), B), 
\end{eqnarray}
where $I'(K)$ is an $N\times (N-M)$ matrix such that 
$I'(K)_{kk} = 1$ for $k\in K$ and $=0$ otherwise. 
Note that $T([N];K\cup\{N+1,\dots,N+M\})$ is given simply by removing
the $M$ columns from the $\mathbbm{1}_N$ block of $T$. 
The determinant of $T([N];K\cup\{N+1,\dots,N+M\})$ is given by
\begin{eqnarray}
\det T([N];K\cup\{N+1,\dots,N+M\})
&=&
{\rm sgn}(K)\det
\left(\begin{array}{ccccccc}
1 &   &        &    & \varphi_1(k_1) & \dots & \varphi_M(k_1) \\
  & 1 &        &    & \varphi_1(k_2) & \dots & \varphi_M(k_2) \\
  &   & \ddots &    & \vdots         &       & \vdots         \\
  &   &        &  1 & \varphi_1(k_{N-M}) & \dots & \varphi_M(k_{N-M})\\
0 &   &        &    & \varphi_1(k_1^c) & \dots & \varphi_M(k_1^c) \\
  & 0 &        &    & \varphi_1(k_2^c) & \dots & \varphi_M(k_2^c) \\
  &   & \ddots &    & \vdots      &       & \vdots         \\
  &   &        &  0 & \varphi_1(k_{M}^c) & \dots & \varphi_M(k_{M}^c)
\end{array}\right)
\\
&=&
{\rm sgn}(K)\det
\left(\begin{array}{cc}
\mathbbm{1}_{N-M} & B(K,[M])\\
O_{M} & B(K^c,[M])\\
\end{array}\right), 
\end{eqnarray}
where $O_M$ is the $M\times M$ zero matrix.
The ${\rm sgn}(K)$ arises from the rearrangement of the rows. 
Thus one obtains
\begin{eqnarray}
\det T([N];K\cup\{N+1,\dots,N+M\}) = {\rm sgn}(K)\det \mathbbm{1}_{N-M}\det B(K^c;[M])
={\rm sgn}(K)\det B(K^c;[M]). 
\label{eq:detT}
\end{eqnarray}
Substituting Eq. \eqref{eq:detT} into Eq. \eqref{eq:bcsx2} leads to
\begin{eqnarray}
&&
\sum_{\sigma\in S_N}{\rm sgn}(\sigma)
\varphi_1(x_{\sigma(1)})\dots\varphi_M(x_{\sigma(M)})
Z(x_{\sigma(M+1)},x_{\sigma(M+2)})\dots Z(x_{\sigma(N-1)},x_{\sigma(N)})
\nonumber\\
&\propto&
\sum_{K\in \left(\begin{array}{c}[N]\\N-M\end{array}\right)}
{\rm pf}{\cal Z}(K;K)\det T([N];K\cup\{N+1,\dots,N+M\}).
\end{eqnarray}
Now the wave function takes the form of a sum of products of a Pfaffian and a determinant. 
It follows from the theorem in Refs. \cite{IW95,Oka19} that 
\begin{eqnarray}
\sum_{K\in \left(\begin{array}{c}[N]\\N-M\end{array}\right)}
{\rm pf}{\cal Z}(K;K)\det T([N];K\cup\{N+1,\dots,N+M\}
&=&
(-1)^{M(M-1)/2}{\rm pf}
\left(\begin{array}{cc}
G{\cal Z}G^T & H \\
-H^T & O_M
\end{array}\right), 
\end{eqnarray}
where $G=T([N];\{1,\dots,N\})$ and $H=T([N];\{N+1,\dots,N+M\})$. 
\end{widetext}
In the present case, $G=\mathbbm{1}_N$ and $H=B$, and thus one finally obtains
\begin{align}
\langle x_1,\dots,x_N|\Psi\rangle \propto
(-1)^{M(M-1)/2}
{\rm pf}
\left(\begin{array}{cc}
{\cal Z} & B \\
-B^T & O_M
\end{array}\right). 
\label{eq:bcsxfin}
\end{align}
As mentioned earlier, this formula is valid also when $M={\rm odd}$ and $N-M = {\rm even}$ \cite{IW95}. Therefore it is applicable to an odd nuclei in which 
there is a blocked level. 
Notice that, in the HF case where $N=M$, Eq. \eqref{eq:bcsxfin} reduces to 
the Slater determinant \cite{IW95,Ba08,Oka19}:
\begin{eqnarray}
\Psi(x_1,x_2,...x_N) 
&\propto&
(-1)^{M(M-1)/2}
{\rm pf}\left(
\begin{array}{cc}
O_N & B \\
-B^T & O_N \\
\end{array}
\right)
\nonumber\\
&=&
\det B.
\end{eqnarray}
Therefore, Eq. \eqref{eq:bcsxfin} for the BCS-type wave functions 
includes the Slater determinant for the HF case as a special case.

\end{document}